\documentclass[a4paper,12pt]{article}
\usepackage{epsfig}
\usepackage{amssymb}
\usepackage{amsfonts}
\usepackage{amsmath}
\usepackage{euscript}
\usepackage{verbatim}
\usepackage{latexsym}
\usepackage{graphicx}

	\usepackage{graphicx}	
	\usepackage{subfigure}
	\usepackage{wrapfig}
	\usepackage[T1]{fontenc}
	\usepackage{mathtools}
	\usepackage{float}
	
	\newcommand{\fsl}[1]{\ensuremath{\mathrlap{\!\not{\phantom{#1}}}#1}}
	\newcommand{\overbar}[1]{\mkern 1.5mu\overline{\mkern-1.5mu#1\mkern-1.5mu}\mkern 1.5mu}

	\newcommand{\bpsi}{\overbar{\psi}}
	\newcommand{\bV}{\overbar{V}}
	\newcommand{\bphi}{\overbar{\phi}}

\usepackage{tikz}
\usetikzlibrary{shapes.geometric, arrows,patterns,snakes}
\tikzstyle{ellip} = [ellipse, minimum width=3cm, minimum height=1cm,text centered, draw=black]

\newskip\humongous \humongous=0pt plus 1000pt minus 1000pt

\newif\ifdtup

\catcode`\@=11

\@addtoreset{equation}{section}

\def\@normalsize{\@setsize\normalsize{15pt}\xiipt\@xiipt
\abovedisplayskip 14pt plus3pt minus3pt%
\belowdisplayskip \abovedisplayskip
\abovedisplayshortskip \z@ plus3pt%
\belowdisplayshortskip 7pt plus3.5pt minus0pt}

\def\small{\@setsize\small{13.6pt}\xipt\@xipt
\abovedisplayskip 13pt plus3pt minus3pt%
\belowdisplayskip \abovedisplayskip
\abovedisplayshortskip \z@ plus3pt%
\belowdisplayshortskip 7pt plus3.5pt minus0pt
\def\@listi{\parsep 4.5pt plus 2pt minus 1pt
     \itemsep \parsep
     \topsep 9pt plus 3pt minus 3pt}}

\relax

\catcode`@=12

\catcode`\@=11

\def\section{\@startsection{section}{1}{\z@}{3.5ex plus 1ex minus
   .2ex}{2.3ex plus .2ex}{\large\bf}}

\def\SymBoxes#1#2#3#4{\newdimen\un@t \un@t#3%
\raisebox{#1}{\rule{#2\un@t}{#4}\hskip-#2\un@t
\@tempdimb\un@t \advance\@tempdimb by-#4\@tempcntb#2\relax%
\@whilenum{\@tempcntb>0}\do{
\rule{#4}{\un@t}\hskip\@tempdimb \advance\@tempcntb by\m@ne}%
\hskip-#2\un@t \rule[\un@t]{#2\un@t}{#4}%
\rule[\un@t]{#4}{#4}\hskip-#4
\rule{#4}{\un@t}}\hskip-#4}                

\begin{document}

\newcommand{\beq}{\begin{equation}}
\newcommand{\eeq}{\end{equation}}
\newcommand{\bea}{\begin{eqnarray}}
\newcommand{\eea}{\end{eqnarray}}
\newcommand{\beas}{\begin{eqnarray*}}
\newcommand{\eeas}{\end{eqnarray*}}
\newcommand{\defi}{\stackrel{\rm def}{=}}
\newcommand{\non}{\nonumber}
\newcommand{\bquo}{\begin{quote}}
\newcommand{\enqu}{\end{quote}}
\renewcommand{\(}{\begin{equation}}
\renewcommand{\)}{\end{equation}}
\def \eqn#1#2{\begin{equation}#2\label{#1}\end{equation}}

\def\e{\epsilon}
\def\IZ{{\mathbb Z}}
\def\IR{{\mathbb R}}
\def\IC{{\mathbb C}}
\def\IQ{{\mathbb Q}}
\def\de{\partial}
\def\Tr{ \hbox{\rm Tr}}
\def\H{ \hbox{\rm H}}
\def\HE{ \hbox{$\rm H^{even}$}}
\def\HO{ \hbox{$\rm H^{odd}$}}
\def\K{ \hbox{\rm K}}
\def\Im{ \hbox{\rm Im}}
\def\Ker{ \hbox{\rm Ker}}
\def\const{\hbox {\rm const.}}
\def\o{\over}
\def\im{\hbox{\rm Im}}
\def\re{\hbox{\rm Re}}
\def\bra{\langle}\def\ket{\rangle}
\def\Arg{\hbox {\rm Arg}}
\def\Re{\hbox {\rm Re}}
\def\Im{\hbox {\rm Im}}
\def\exo{\hbox {\rm exp}}
\def\diag{\hbox{\rm diag}}
\def\longvert{{\rule[-2mm]{0.1mm}{7mm}}\,}
\def\a{\alpha}
\def\dag{{}^{\dagger}}
\def\tq{{\widetilde q}}
\def\p{{}^{\prime}}
\def\W{W}
\def\N{{\cal N}}
\def\hsp{,\hspace{.7cm}}

\def\br{\nonumber\\}
\def\IZ{{\mathbb Z}}
\def\IR{{\mathbb R}}
\def\IC{{\mathbb C}}
\def\IQ{{\mathbb Q}}
\def\IP{{\mathbb P}}
\def \eqn#1#2{\begin{equation}#2\label{#1}\end{equation}}

\newcommand{\C}{\ensuremath{\mathbb C}}
\newcommand{\Z}{\ensuremath{\mathbb Z}}
\newcommand{\R}{\ensuremath{\mathbb R}}
\newcommand{\rp}{\ensuremath{\mathbb {RP}}}
\newcommand{\cp}{\ensuremath{\mathbb {CP}}}
\newcommand{\vac}{\ensuremath{|0\rangle}}
\newcommand{\vact}{\ensuremath{|00\rangle}                    }
\newcommand{\oc}{\ensuremath{\overline{c}}}
\begin{titlepage}
\begin{flushright}
CHEP XXXXX
\end{flushright}
\bigskip
\def\thefootnote{\fnsymbol{footnote}}

\begin{center}
{\Large
{\bf $\epsilon$-Expansion in the Gross-Neveu CFT \\ \vspace{0.1in} 
}
}
\end{center}

\bigskip
\begin{center}
{
Avinash RAJU$^a$\footnote{\texttt{avinash@cts.iisc.ernet.in}} }
\vspace{0.1in}

\end{center}

\renewcommand{\thefootnote}{\arabic{footnote}}

\begin{center}

$^a$ {Center for High Energy Physics,\\
Indian Institute of Science, Bangalore 560012, India}\\

\end{center}

\noindent
\begin{center} {\bf Abstract} \end{center}

We use the recently developed CFT techniques of Rychkov and Tan to compute anomalous dimensions in the $O(N)$ Gross-Neveu model in $d=2+\epsilon$ dimensions. To do this, we extend the ``cowpie contraction" algorithm of arXiv:1506.06616 to theories with fermions. Our results match perfectly with Feynman diagram computations. 


\vspace{1.6 cm}
\vfill

\end{titlepage}

\setcounter{footnote}{0}

\section{Introduction}


In recent work, Rychkov and Tan \cite{Rychkov} have shown that the power of conformal invariance can be used to compute $\epsilon$-expansions at the Wilson-Fisher fixed point (see also \cite{Chethan-Pallab}). This approach is not reliant on Feynman diagrams (and in that sense is non-perturbative\footnote{This should be taken with a pinch of salt -- the epsilon expansion is afterall perturbative. The idea here is that the perturbative parameter in the present approach is not (at least manifestly) the coupling constant.}), and uses only conformal symmetry and analyticity in $\epsilon$ as inputs.

The results of Rychkov-Tan were generalized to other dimensions and other fixed point theories in \cite{Chethan-Pallab}. The computations require a systematic approach to handling contractions of fields in these theories, and a systematic approach for doing this was developed for scalar $O(N)$ theories \cite{Chethan-Pallab}. One of the goals of this paper is to generalize this to CFTs with fermions.

Concretely, we will work with $O(N)$ Gross-Neveu model in $d=2+\epsilon$ dimensions \cite{GN} \footnote{The multiplicative renormalizability of Gross-Neveu model in $2+\e$ dimensions is discussed in \cite{Vasiliev1,Vasiliev2}, our results are unchanged for the $U(N)$ model as well.}. This theory is interesting for various reasons: there is a huge literature on this theory, and its large-N expansion and asymptotic freedom (among various other features) have been thoroughly investigated in the last decades. We generalize the approach of \cite{Rychkov,Chethan-Pallab} to this theory, and verify that the results agree with existing perturbative results in the literature, where they overlap.

The paper is organized as follows. In section 2 we introduce the Gross-Neveu as a Wislon-Fisher CFT along with the axioms that help us along in the computation. In section 3 we give a recursive algorithm, based on \cite{Chethan-Pallab}, to compute OPE coefficients in the free theory and in section 4 we show how these results are matched with that of interacting theory in the $\e \rightarrow 0$ limit which help us determine anomalous dimensions of various composite operators.

Comment added: The paper \cite{rgupta} also discusses the same problem, and even though the details of the algorithm are different, our results agree.

\section{$O(N)$ Gross-Neveu model in $2+\epsilon$ dimensions}\label{sec1}

The Gross-Neveu model action in $d=2+\epsilon$ dimensions is given by

\bea
S = \frac{1}{2\pi}\int d^d \sigma \left[ \bpsi^A \fsl{\partial}\psi^A +  \frac{1}{2}g\mu^{-\epsilon}(\bpsi^A \psi^A)^2 \right]
\eea

\noindent In 2 dimensions, this theory is renormalizable with a dimensionless coupling constant. The coupling constant is proportional to $\e$ and hence this theory describes a weakly coupled fixed point for small values of $\e$. We have introduces a scale $\mu$ to make the coupling constant dimensionless. \\
The engineering dimension of the fields is fixed by the action

\bea
[\psi] \equiv \delta = \frac{1+\e}{2}
\eea

\noindent The equations of motion for this theory are given by

\begin{eqnarray}
\gamma^{\mu}\partial_{\mu}\psi^A + g\mu^{-\epsilon} (\bpsi^B \psi^B)\psi^A &=& 0 \\
\partial_{\mu}\bpsi^A \gamma^{\mu} - g\mu^{-\epsilon} (\bpsi^B \psi^B)\bpsi^A &=& 0 
\end{eqnarray}

\noindent According to \cite{Rychkov}, this equation has to be seen as a conformal multiplet shortening condition, where in the free theory, the operators $(\bpsi^B \psi^B)\psi^A$ and $(\bpsi^B \psi^B)\bpsi^A$ is a primary, but in the interacting theory it is made secondary by above equations. Following \cite{Rychkov}, we formalize the relationship between operators in the free and interacting case by means of following axioms:

\begin{itemize}
\item The interacting theory enjoys conformal symmetry.

\item For any operator in the interacting theory, there is a corresponding operator in the free theory, which the interacting theory operator approaches to in the $\e \rightarrow 0$ limit. \\

\noindent For definiteness, we call the interacting theory operators as $V_{2n}$, $V^{A}_{2n+1\;a}$ and $\overbar{V}^{A}_{2n+1\;a}$\footnote{A word on notations: small latin indices $a$, $b$, $\cdots$ are the spinor indices whereas $A$, $B$, etc stand for $O(N)$ indices} which in the free limit goes to 

\begin{eqnarray}
V_{2n} & \rightarrow & (\bpsi^A \psi^A)^{n}  \\ \nonumber
V^{A}_{2n+1\;a} & \rightarrow & (\bpsi^B \psi^B)^{n} \psi^{A}_{a} \\ \nonumber
\overbar{V}^{A}_{2n+1\;a} & \rightarrow & (\bpsi^B \psi^B)^{n} \bpsi^{A}_{a} \nonumber
\end{eqnarray}

\item Operators $V^{A}_{3\;a}$ and $\overbar{V}^{A}_{3\;a}$ are not primaries, instead they are related to the primaries by the multiplet shortening conditions

\begin{eqnarray}\label{multiplet_short}
\gamma^{\mu}\partial_{\mu}\psi^A = - \alpha(\e) (\bpsi^B \psi^B)\psi^A  \\ \nonumber
\partial_{\mu}\bpsi^A \gamma^{\mu} = \alpha(\e) (\bpsi^B \psi^B)\bpsi^A
\end{eqnarray}

\noindent This puts restrictions on the dimensions of these operators

\bea\label{dimconstraint}
\Delta_{3} = \Delta_{1} + 1
\eea

\noindent The proportionality constant $\alpha(\e)$ can be fixed later using the axioms above. All other operators $V_{m}$, $m \neq 3$, are primaries. 

\end{itemize}

\noindent The two-point function of two primaries of same dimension $\Delta_1$ is 

\bea
\langle V^{A}_{1\;a}(x_1) \overbar{V}^{B}_{1\;b}(x_2) \rangle = \frac{(\fsl{x}_{12})_{ab}}{(x_{12}^{2})^{\Delta_1 + \frac{1}{2}}}\delta^{AB} \label{int2pt}
\eea

\noindent In the free limit this becomes
\bea
\langle \psi^{A}_{a}(x_1) \bpsi^{B}_{b}(x_2) \rangle = \frac{(\fsl{x}_{12})_{ab}}{x_{12}^{2}}\delta^{AB} \label{free2pt}
\eea

\noindent The anomalous dimension is defined as the difference between the actual scaling dimension of the operator and the engineering dimension, i.e, $\Delta_{n} = n\delta + \gamma_{n}$. We also make the crucial assumption that the anomalous dimensions are analytic functions of $\e$ and therefore admits a power series expansion

\bea
\gamma_n=y_{n,1} \e + y_{n,2} \e^2+ ...
\eea

\noindent Our first task is to fix $\alpha$ in \eqref{multiplet_short}. Differentiating \eqref{int2pt} and substituting appropriate factors of $\gamma$ matrices, we obtain

\begin{eqnarray}\label{diff_correlator}
(\gamma^{\mu})_{ca}\langle \partial_{1\; \mu}V^{A}_{1\;a}(x_1) \partial_{2\; \nu}\bV^{B}_{1\; b}(x_2) \rangle (\gamma^{\nu})_{bd} &=& (\gamma^{\mu})_{ca}\partial_{1\; \mu} \partial_{2\; \nu} \left( \frac{(\fsl{x}_{12})_{ab}}{(x_{12}^2)^{\Delta_1 + \frac{1}{2}}} \right) (\gamma^{\nu})_{bd}\delta^{AB} \\ \nonumber &=& -(2\Delta_1 + 1)(2\Delta_1 +1 -d) \frac{(\fsl{x}_{12})_{cd}}{(x^{2}_{12})^{\Delta_1 + \frac{3}{2}}} \delta^{AB}
\end{eqnarray}

\noindent Left hand side of \eqref{diff_correlator} takes the form

\begin{equation}
-\alpha^2 \langle V^{A}_{3\;c}(x) \bV^{B}_{3\;d}(y) \rangle 
\end{equation}

\noindent which in the free limit evaluates to

\bea
-\alpha^{2}(\e)(N-1)\frac{(\fsl{x}_{12})_{cd}}{(x^{2}_{12})^{2}}\delta^{AB}
\eea

\noindent Comparing both sides, we obtain

\begin{equation}\label{alpha}
\alpha = \sigma \sqrt{\frac{4\gamma_1}{N-1}}
\end{equation}

\noindent where $\sigma=\pm 1$. The exact sign will be determined later. Following \cite{Rychkov, Chethan-Pallab}, we consider correlators of the form

\bea
\langle V_{2n}(x_1) V^{A}_{2n+1\;a}(x_2) \bV^{B}_{1\;b}(x_3) \rangle, \quad \langle V_{2n}(x_1) V^{A}_{2n+1\;a}(x_2) \bV^{B}_{3\;b}(x_3) \rangle
\eea

\noindent which in the free limit goes to

\bea
\langle \phi_{2n}(x_1) \phi_{2n+1\; a}^{A}(x_2) \bphi^{B}_{1\;b}(x_3) \rangle, \quad  \langle \phi_{2n}(x_1) \phi_{2n+1\; a}^{A}(x_2) \bphi^{B}_{3\;b}(x_3)
\eea

\noindent where we have introduced operators $\phi_{2n}$ and $\phi_{2n+1\; a}^{A}$ as a shorthand for $(\bpsi^{B}_{b}\psi^{B}_{b})^n$ and $(\bpsi^{B}_{b} \psi^{B}_{b})^n \psi^{A}_{a}$. The reason we are interested in these correlators is because of its sensitivity to multiplet recombination. To see this, we notice that in the free theory, $\phi_{2n} \times \phi_{2n+1\; a}^{A}$ OPE contains operators $\psi^{A}_{a}$ and $(\bpsi^{B}_{b}\psi^{b}_{b})\psi^{A}_{a}$ whereas in the interacting theory $V_{2n} \times V^{A}_{2n+1\;a}$ OPE only contains $V_1$ as the primary. The coefficients in both cases are independently computable and by Axiom:2, we expect them to match in the limit $\e \rightarrow 0$.

\noindent In the free case, we have following OPE

\bea\label{freeOPE1}
\phi_{2n}(x_1) \times \phi_{2n+1\;a}^{A}(x_2) \supset f_{2n} (x^{2}_{12})^{-n}\left(\psi_{a}^{A} + \rho_{2n} (\fsl{x}_{12})_{ab} (\bpsi \psi)\psi^{A}_{b} \right)
\eea

\noindent The coefficients $f_{2n}$ and $\rho_{2n}$ can be determined by counting the number of Wick contractions. In next section, we provide an algorithm, based on \cite{Chethan-Pallab}, to determine these coefficients for arbitrary $n$. This is matched with the interacting theory OPE

\begin{eqnarray}\label{intOPE}
V_{2n}(x_1)\times V^{A}_{2n+1\;a}(x_2)  & \supset & \tilde{f}_{2n} (x^{2}_{12})^{-\frac{1}{2}[\Delta_{2n} + \Delta_{2n+1}-\Delta_1]} \\ \nonumber &\;& \Big[ \delta_{ac} + q_1 \delta_{ac} x_{12}^{\mu} \partial_{2\; \mu} + q_2 (\fsl{x_{12}}\fsl{\partial}_{2})_{ac}  \Big] V^{A}_{1\;c}(x_2)
\end{eqnarray}
  
\section{Counting contractions}\label{cowpie}

\noindent We now turn our attention to computing $f$ and $\rho$ coefficients in \eqref{freeOPE1}. Apart from \eqref{freeOPE1}, we also need OPE's of the form

\bea\label{freeOPE2}
\phi^{A}_{2n+1\;a}(x_1) \times \phi_{2n+2}(x_2) \supset f_{2n+1} (x^{2}_{12})^{-(n+1)}\Big[(\fsl{x}_{12})_{ab}\psi_{b}^{A} + \rho_{2n+1} x^{2}_{12} (\bpsi \psi)\psi^{A}_{a} \Big]
\eea

\noindent which are used to fix the anomalous dimensions of odd operators. In \cite{Chethan-Pallab} a recursive algorithm was used to count Wick contractions, which can be adapted for the fermions. The Wick contractions can then be viewed as various ways of connecting upper and lower rows, resulting in recursive equations. In the case of fermions, the principle is essentially the same, but the contractions have a bit more structure. We use '$+$' and '$-$' to denote $\bpsi$ and $\psi$ respectively. To capture the contractions, we introduce the quantity $F^{p,r_{+},r_{-}}_{p+q,s_{+},s_{-};m_{+},m_{-}}$, where $p$ is the number of upper double cow-pies which stand for $\bpsi \psi$, $r_{+}$ is the number of upper single cow-pies of '$+$' type, $r_{-}$ is the number of upper single cow-pies of '$+$' type, $p+q$ is the number of lower double cow-pies, $s_{\pm}$ is the number of lower single cow-pies of type $\pm$, $m_{\pm}$ is the number of uncontracted $\bpsi$s and  $\psi$s respectively. A contraction is always between an upper $+$ and a lower $-$ or vice-versa.

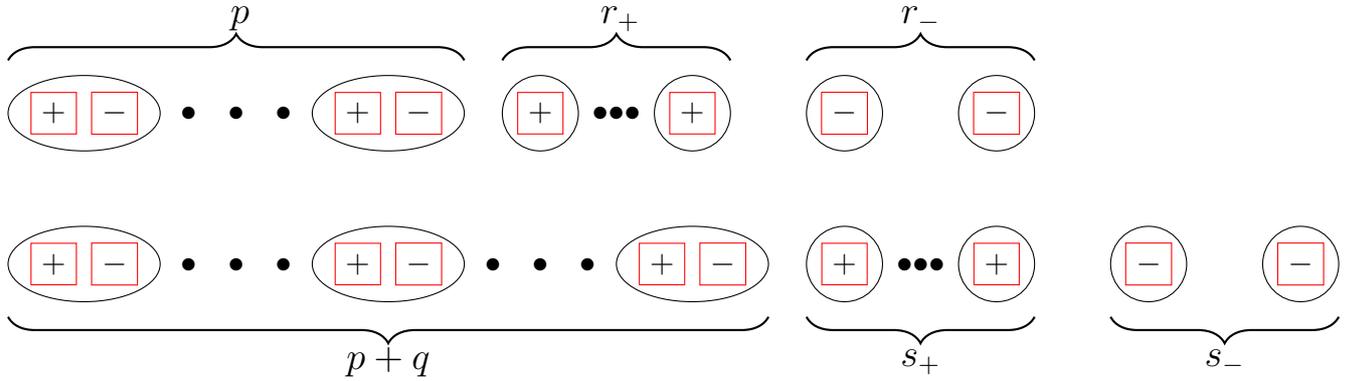
\begin{figure}[H]
\begin{tikzpicture}

\draw (2,-0.25) node[anchor=south] {\textbullet\quad\textbullet\quad\textbullet};
\draw (7,-0.25) node[anchor=south] {\textbullet\textbullet\textbullet};
\begin{scope}
    \draw (0,0) ellipse (1cm and 0.5cm);
    \node[draw=red] at (-0.4,0)  {$+$};
    \node[draw=red] at (0.4,0)  {$-$} ;
\end{scope}

\begin{scope}[xshift=4cm]
\draw (0,0) ellipse (1cm and 0.5cm);
\node[draw=red] at (-0.4,0)  {$+$};
\node[draw=red] at (0.4,0)  {$-$} ;
\end{scope}

\begin{scope}[xshift=6cm]
\draw (0,0) circle (0.5cm);
\node[draw=red] at (0,0)  {$+$} ;
\end{scope}

\begin{scope}[xshift=8cm]
\draw (0,0) circle (0.5cm);
\node[draw=red] at (0,0)  {$+$} ;
\end{scope}
\begin{scope}[xshift=10cm]
\draw (0,0) circle (0.5cm);
\node[draw=red] at (0,0)  {$-$} ;
\end{scope}

\begin{scope}[xshift=12cm]
\draw (0,0) circle (0.5cm);
\node[draw=red] at (0,0)  {$-$} ;
\end{scope}
\draw (2,-2.25) node[anchor=south] {\textbullet\quad\textbullet\quad\textbullet};
\draw (6,-2.25) node[anchor=south] {\textbullet\quad\textbullet\quad\textbullet};
\draw (11,-2.25) node[anchor=south] {\textbullet\textbullet\textbullet};

\begin{scope}[yshift=-2cm]
\draw (0,0) ellipse (1cm and 0.5cm);
\node[draw=red] at (-0.4,0)  {$+$};
\node[draw=red] at (0.4,0)  {$-$} ;
\end{scope}

\begin{scope}[xshift=4cm,yshift=-2cm]
\draw (0,0) ellipse (1cm and 0.5cm);
\node[draw=red] at (-0.4,0)  {$+$};
\node[draw=red] at (0.4,0)  {$-$} ;
\end{scope}
 
\begin{scope}[xshift=8cm,yshift=-2cm]
\draw (0,0) ellipse (1cm and 0.5cm);
\node[draw=red] at (-0.4,0)  {$+$};
\node[draw=red] at (0.4,0)  {$-$} ;
\end{scope}

\begin{scope}[xshift=10cm,yshift=-2cm]
\draw (0,0) circle (0.5cm);
\node[draw=red] at (0,0)  {$+$} ;
\end{scope}

\begin{scope}[xshift=12cm,yshift=-2cm]
\draw (0,0) circle (0.5cm);
\node[draw=red] at (0,0)  {$+$} ;
\end{scope}

\begin{scope}[xshift=14cm,yshift=-2cm]
\draw (0,0) circle (0.5cm);
\node[draw=red] at (0,0)  {$-$} ;
\end{scope}

\begin{scope}[xshift=16cm,yshift=-2cm]
\draw (0,0) circle (0.5cm);
\node[draw=red] at (0,0)  {$-$} ;
\end{scope}

\begin{scope}[yshift=0.7cm]
\draw [thick,decorate,decoration={brace,amplitude=10pt}] (-1.0,0) -- (5.0,0);
\node[] at (2.05,0.55) {\large $p$}; 
\end{scope}

\begin{scope}[yshift=0.7cm]
\draw [thick,decorate,decoration={brace,amplitude=10pt}] (5.5,0) -- (8.5,0);
\node[] at (7.05,0.55) {\large $r_{+}$}; 
\end{scope}

\begin{scope}[yshift=0.7cm]
\draw [thick,decorate,decoration={brace,amplitude=10pt}] (9.5,0) -- (12.5,0);
\node[] at (11.0,0.55) {\large $r_{-}$}; 
\end{scope}

\begin{scope}[yshift=-2.7cm]
\draw [thick,decorate,decoration={brace,mirror,amplitude=10pt}] (-1.0,0) -- (9.0,0);
\node[] at (4.0,-.6) {\large $p+q$}; 
\end{scope}

\begin{scope}[yshift=-2.7cm]
\draw [thick,decorate,decoration={brace,mirror,amplitude=10pt}] (9.5,0) -- (12.5,0);
\node[] at (11.0,-.6) {\large $s_{+}$}; 
\end{scope}

\begin{scope}[yshift=-2.7cm]
\draw [thick,decorate,decoration={brace,mirror,amplitude=10pt}] (13.5,0) -- (16.5,0);
\node[] at (15.0,-.6) {\large $s_{-}$}; 
\end{scope}

\end{tikzpicture}
\end{figure}

\noindent The various coefficients $f$s and $\rho$s in our notation becomes

\begin{eqnarray}
f_{2p} = F^{p,0,0}_{p,0,1;0,1} & \quad & f_{2p}\rho_{2p} = F^{p,0,0}_{p,0,1;1,2} \\ \nonumber
f_{2p+1} = F^{p,0,1}_{p+1,0,0;0,1} & \quad & f_{2p+1}\rho_{2p+1} = F^{p,0,0}_{p+1,0,0;1,2} \nonumber
\end{eqnarray}

\subsection{$\mathbf{f_{2p}}$}

\noindent There are 3 different kind of contractions that are possible. Of the first type, the two kernels of the $p^{\mathrm{th}}$ double cow-pie are contracted with two kernels of same lower cow-pie. This gives a factor of $Np$. The second possibility is to contract the two kernels of upper cow-pie to two different kernels of lower double cow-pie resulting in a factor of $-p(p-1)$ and the last possibility is to contract one of the kernels of upper double cow-pie with a kernel in lower double cow-pie and the second kernel of upper cow-pie with the single kernel of lower row. This gives a factor of $-p$. So, the resulting contraction can now be expressed as following recursion equation

\begin{equation}
F^{p,0,0}_{p,0,1;0,1} = (Np - p(p-1) - p)F^{p-1;0}_{p-1,1;0,1}
\end{equation}

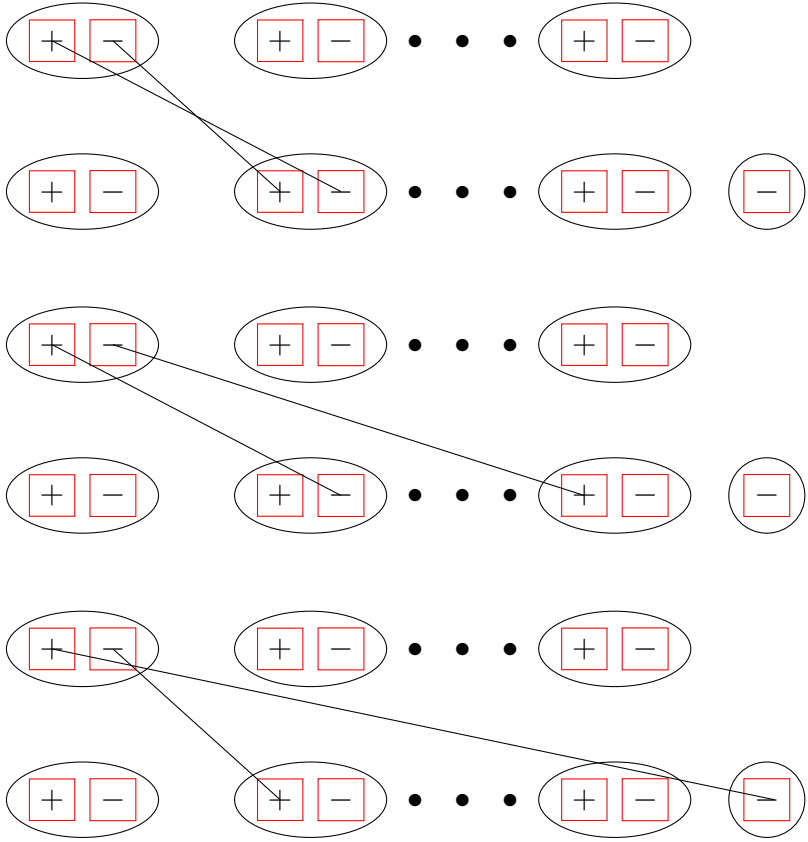
\begin{figure}[H]
\begin{tikzpicture}

\draw (5,-0.25) node[anchor=south] {\textbullet\quad\textbullet\quad\textbullet};

\begin{scope}
    \draw (0,0) ellipse (1cm and 0.5cm);
    \node[draw=red] at (-0.4,0)  {$+$};
    \node[draw=red] at (0.4,0)  {$-$} ;
\end{scope}

\begin{scope}[xshift=3cm]
\draw (0,0) ellipse (1cm and 0.5cm);
\node[draw=red] at (-0.4,0)  {$+$};
\node[draw=red] at (0.4,0)  {$-$} ;
\end{scope}
 
\begin{scope}[xshift=7cm]
\draw (0,0) ellipse (1cm and 0.5cm);
\node[draw=red] at (-0.4,0)  {$+$};
\node[draw=red] at (0.4,0)  {$-$} ;
\end{scope}

\draw (5,-2.25) node[anchor=south] {\textbullet\quad\textbullet\quad\textbullet};

\begin{scope}[yshift=-2cm]
\draw (0,0) ellipse (1cm and 0.5cm);
\node[draw=red] at (-0.4,0)  {$+$};
\node[draw=red] at (0.4,0)  {$-$} ;
\end{scope}

\begin{scope}[xshift=3cm,yshift=-2cm]
\draw (0,0) ellipse (1cm and 0.5cm);
\node[draw=red] at (-0.4,0)  {$+$};
\node[draw=red] at (0.4,0)  {$-$} ;
\end{scope}
 
\begin{scope}[xshift=7cm,yshift=-2cm]
\draw (0,0) ellipse (1cm and 0.5cm);
\node[draw=red] at (-0.4,0)  {$+$};
\node[draw=red] at (0.4,0)  {$-$} ;
\end{scope}

\begin{scope}[xshift=9cm,yshift=-2cm]
\draw (0,0) circle (0.5cm);
\node[draw=red] at (0,0)  {$-$} ;

\end{scope}

\draw (-0.4,0) -- (3.4,-2);
\draw (0.4,0) --  (2.6,-2);


\end{tikzpicture}
\end{figure}

\begin{figure}[H]
\begin{tikzpicture}

\draw (5,-0.25) node[anchor=south] {\textbullet\quad\textbullet\quad\textbullet};

\begin{scope}
    \draw (0,0) ellipse (1cm and 0.5cm);
    \node[draw=red] at (-0.4,0)  {$+$};
    \node[draw=red] at (0.4,0)  {$-$} ;
\end{scope}

\begin{scope}[xshift=3cm]
\draw (0,0) ellipse (1cm and 0.5cm);
\node[draw=red] at (-0.4,0)  {$+$};
\node[draw=red] at (0.4,0)  {$-$} ;
\end{scope}
 
\begin{scope}[xshift=7cm]
\draw (0,0) ellipse (1cm and 0.5cm);
\node[draw=red] at (-0.4,0)  {$+$};
\node[draw=red] at (0.4,0)  {$-$} ;
\end{scope}

\draw (5,-2.25) node[anchor=south] {\textbullet\quad\textbullet\quad\textbullet};

\begin{scope}[yshift=-2cm]
\draw (0,0) ellipse (1cm and 0.5cm);
\node[draw=red] at (-0.4,0)  {$+$};
\node[draw=red] at (0.4,0)  {$-$} ;
\end{scope}

\begin{scope}[xshift=3cm,yshift=-2cm]
\draw (0,0) ellipse (1cm and 0.5cm);
\node[draw=red] at (-0.4,0)  {$+$};
\node[draw=red] at (0.4,0)  {$-$} ;
\end{scope}
 
\begin{scope}[xshift=7cm,yshift=-2cm]
\draw (0,0) ellipse (1cm and 0.5cm);
\node[draw=red] at (-0.4,0)  {$+$};
\node[draw=red] at (0.4,0)  {$-$} ;
\end{scope}

\begin{scope}[xshift=9cm,yshift=-2cm]
\draw (0,0) circle (0.5cm);
\node[draw=red] at (0,0)  {$-$} ;

\end{scope}

\draw (-0.4,0) -- (3.4,-2);
\draw (0.4,0) --  (6.6,-2);


\end{tikzpicture}
\end{figure}

\begin{figure}[H]
\begin{tikzpicture}

\draw (5,-0.25) node[anchor=south] {\textbullet\quad\textbullet\quad\textbullet};

\begin{scope}
    \draw (0,0) ellipse (1cm and 0.5cm);
    \node[draw=red] at (-0.4,0)  {$+$};
    \node[draw=red] at (0.4,0)  {$-$} ;
\end{scope}

\begin{scope}[xshift=3cm]
\draw (0,0) ellipse (1cm and 0.5cm);
\node[draw=red] at (-0.4,0)  {$+$};
\node[draw=red] at (0.4,0)  {$-$} ;
\end{scope}
 
\begin{scope}[xshift=7cm]
\draw (0,0) ellipse (1cm and 0.5cm);
\node[draw=red] at (-0.4,0)  {$+$};
\node[draw=red] at (0.4,0)  {$-$} ;
\end{scope}

\draw (5,-2.25) node[anchor=south] {\textbullet\quad\textbullet\quad\textbullet};

\begin{scope}[yshift=-2cm]
\draw (0,0) ellipse (1cm and 0.5cm);
\node[draw=red] at (-0.4,0)  {$+$};
\node[draw=red] at (0.4,0)  {$-$} ;
\end{scope}

\begin{scope}[xshift=3cm,yshift=-2cm]
\draw (0,0) ellipse (1cm and 0.5cm);
\node[draw=red] at (-0.4,0)  {$+$};
\node[draw=red] at (0.4,0)  {$-$} ;
\end{scope}
 
\begin{scope}[xshift=7cm,yshift=-2cm]
\draw (0,0) ellipse (1cm and 0.5cm);
\node[draw=red] at (-0.4,0)  {$+$};
\node[draw=red] at (0.4,0)  {$-$} ;
\end{scope}

\begin{scope}[xshift=9cm,yshift=-2cm]
\draw (0,0) circle (0.5cm);
\node[draw=red] at (0,0)  {$-$} ;

\end{scope}

\draw (-0.4,0) -- (9.1,-2);
\draw (0.4,0) --  (2.6,-2);


\end{tikzpicture}
\end{figure}

\noindent This recursion equation can be solved along with the launching condition $F^{0;0}_{0,1;0,1}=1$ and we obtain

\begin{equation}\label{f2p}
f_{2p} = p! (N-1)(N-2)\cdots (N-p)
\end{equation}

\subsection{$\mathbf{f_{2p+1}}$}

\noindent There are only two types of contractions possible, analogous to the first two types above. The recursion equation can therefore be written by inspection

\begin{equation}
F^{p;1}_{p+1,0;0,1} = (N(p+1) - p(p+1))F^{p-1,0,1}_{p,0,0;0,1}
\end{equation}

\noindent with the lauching condition $F^{0;1}_{1,0;0,1}=1$ which gives

\begin{equation}\label{f2pp1}
f_{2p+1} = (p+1)!(N-1)(N-2)\cdots (N-p)
\end{equation}

\subsection{$\mathbf{f_{2p}\rho_{2p}}$}

\noindent Here the first possibility involves contracting both the kernels of upper double cow-pie with lower cow-pies analogous to the computation of $f_{2p}$. This gives a factor of $Np - p(p-1) - p$. Another possibility involves contracting '$+$' of an upper double cow-pie with the single '$-$' in the lower row. This gives a factor of $-F^{p-1;0}_{p-1,0;1,1}$.

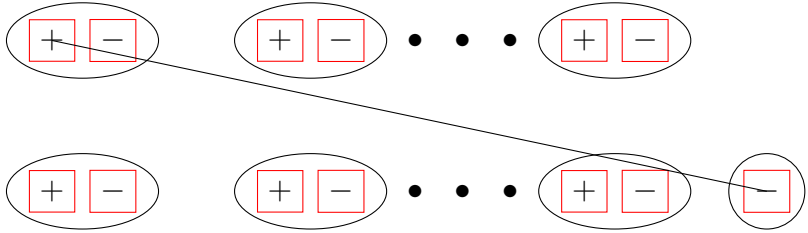
\begin{figure}[H]
\begin{tikzpicture}

\draw (5,-0.25) node[anchor=south] {\textbullet\quad\textbullet\quad\textbullet};

\begin{scope}
    \draw (0,0) ellipse (1cm and 0.5cm);
    \node[draw=red] at (-0.4,0)  {$+$};
    \node[draw=red] at (0.4,0)  {$-$} ;
\end{scope}

\begin{scope}[xshift=3cm]
\draw (0,0) ellipse (1cm and 0.5cm);
\node[draw=red] at (-0.4,0)  {$+$};
\node[draw=red] at (0.4,0)  {$-$} ;
\end{scope}
 
\begin{scope}[xshift=7cm]
\draw (0,0) ellipse (1cm and 0.5cm);
\node[draw=red] at (-0.4,0)  {$+$};
\node[draw=red] at (0.4,0)  {$-$} ;
\end{scope}

\draw (5,-2.25) node[anchor=south] {\textbullet\quad\textbullet\quad\textbullet};

\begin{scope}[yshift=-2cm]
\draw (0,0) ellipse (1cm and 0.5cm);
\node[draw=red] at (-0.4,0)  {$+$};
\node[draw=red] at (0.4,0)  {$-$} ;
\end{scope}

\begin{scope}[xshift=3cm,yshift=-2cm]
\draw (0,0) ellipse (1cm and 0.5cm);
\node[draw=red] at (-0.4,0)  {$+$};
\node[draw=red] at (0.4,0)  {$-$} ;
\end{scope}
 
\begin{scope}[xshift=7cm,yshift=-2cm]
\draw (0,0) ellipse (1cm and 0.5cm);
\node[draw=red] at (-0.4,0)  {$+$};
\node[draw=red] at (0.4,0)  {$-$} ;
\end{scope}

\begin{scope}[xshift=9cm,yshift=-2cm]
\draw (0,0) circle (0.5cm);
\node[draw=red] at (0,0)  {$-$} ;
\end{scope}

\draw (-0.4,0) -- (9.0,-2);


\end{tikzpicture}
\end{figure}

\noindent Other possibilities involve single contractions of upper kernels, which can be done in following ways: (a) '$+$' of the upper double cow-pie contracted with a '$-$' of the lower double cow-pie, (b) '$-$' from the upper double cow-pie with a $+$ from lower double cow-pie. One can see by explicit computation for the lower orders (\textit{i.e.} $p=2,3,\cdots$)  that their contribution is given by $-p F^{p-1,0,0}_{p-1,0,1;1,2}$. Notice that the coefficient is different from the naive expectation because not all single contractions are independent and we must be careful to avoid over-counting and to keep track of the index structure.

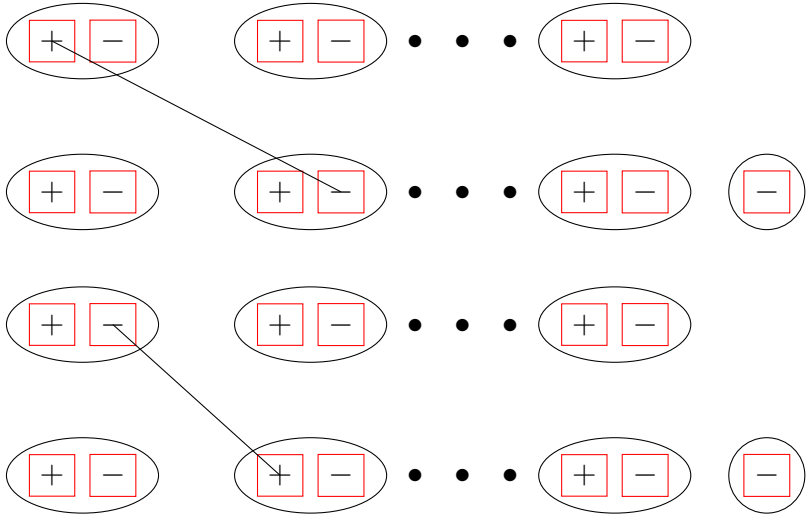
\begin{figure}[H]
\begin{tikzpicture}

\draw (5,-0.25) node[anchor=south] {\textbullet\quad\textbullet\quad\textbullet};

\begin{scope}
    \draw (0,0) ellipse (1cm and 0.5cm);
    \node[draw=red] at (-0.4,0)  {$+$};
    \node[draw=red] at (0.4,0)  {$-$} ;
\end{scope}

\begin{scope}[xshift=3cm]
\draw (0,0) ellipse (1cm and 0.5cm);
\node[draw=red] at (-0.4,0)  {$+$};
\node[draw=red] at (0.4,0)  {$-$} ;
\end{scope}
 
\begin{scope}[xshift=7cm]
\draw (0,0) ellipse (1cm and 0.5cm);
\node[draw=red] at (-0.4,0)  {$+$};
\node[draw=red] at (0.4,0)  {$-$} ;
\end{scope}

\draw (5,-2.25) node[anchor=south] {\textbullet\quad\textbullet\quad\textbullet};

\begin{scope}[yshift=-2cm]
\draw (0,0) ellipse (1cm and 0.5cm);
\node[draw=red] at (-0.4,0)  {$+$};
\node[draw=red] at (0.4,0)  {$-$} ;
\end{scope}

\begin{scope}[xshift=3cm,yshift=-2cm]
\draw (0,0) ellipse (1cm and 0.5cm);
\node[draw=red] at (-0.4,0)  {$+$};
\node[draw=red] at (0.4,0)  {$-$} ;
\end{scope}
 
\begin{scope}[xshift=7cm,yshift=-2cm]
\draw (0,0) ellipse (1cm and 0.5cm);
\node[draw=red] at (-0.4,0)  {$+$};
\node[draw=red] at (0.4,0)  {$-$} ;
\end{scope}

\begin{scope}[xshift=9cm,yshift=-2cm]
\draw (0,0) circle (0.5cm);
\node[draw=red] at (0,0)  {$-$} ;

\end{scope}

\draw (-0.4,0) -- (3.4,-2);


\end{tikzpicture}
\end{figure}

\begin{figure}[H]
\begin{tikzpicture}

\draw (5,-0.25) node[anchor=south] {\textbullet\quad\textbullet\quad\textbullet};

\begin{scope}
    \draw (0,0) ellipse (1cm and 0.5cm);
    \node[draw=red] at (-0.4,0)  {$+$};
    \node[draw=red] at (0.4,0)  {$-$} ;
\end{scope}

\begin{scope}[xshift=3cm]
\draw (0,0) ellipse (1cm and 0.5cm);
\node[draw=red] at (-0.4,0)  {$+$};
\node[draw=red] at (0.4,0)  {$-$} ;
\end{scope}
 
\begin{scope}[xshift=7cm]
\draw (0,0) ellipse (1cm and 0.5cm);
\node[draw=red] at (-0.4,0)  {$+$};
\node[draw=red] at (0.4,0)  {$-$} ;
\end{scope}

\draw (5,-2.25) node[anchor=south] {\textbullet\quad\textbullet\quad\textbullet};

\begin{scope}[yshift=-2cm]
\draw (0,0) ellipse (1cm and 0.5cm);
\node[draw=red] at (-0.4,0)  {$+$};
\node[draw=red] at (0.4,0)  {$-$} ;
\end{scope}

\begin{scope}[xshift=3cm,yshift=-2cm]
\draw (0,0) ellipse (1cm and 0.5cm);
\node[draw=red] at (-0.4,0)  {$+$};
\node[draw=red] at (0.4,0)  {$-$} ;
\end{scope}
 
\begin{scope}[xshift=7cm,yshift=-2cm]
\draw (0,0) ellipse (1cm and 0.5cm);
\node[draw=red] at (-0.4,0)  {$+$};
\node[draw=red] at (0.4,0)  {$-$} ;
\end{scope}

\begin{scope}[xshift=9cm,yshift=-2cm]
\draw (0,0) circle (0.5cm);
\node[draw=red] at (0,0)  {$-$} ;

\end{scope}

\draw (0.4,0) --  (2.6,-2);


\end{tikzpicture}
\end{figure}

\noindent Thus we get the recursion equation

\begin{equation}
F^{p,0,0}_{p,0,1;1,2} = p\left[N - p -1 \right]F^{p-1,0,0}_{p-1,0;1,2} - F^{p-1,0,1}_{p,0,0;1,1}
\end{equation}

\noindent $F^{p-1,0,1}_{p,0,0;1,1}$ can be again evaluated using the cow-pie formalism and its recursion equation is given by

\begin{equation}
F^{p;0}_{p,0;1,1} = (p+1)(N-p)F^{p-1,0}_{p,0;1,1}
\end{equation}

\noindent This system of recursion equations can be solved using the launching condition $F^{0,0,0}_{0,0,1;1,2}=0$ and $F^{0,0,0}_{1,0,0;1,1}=1$. Using the expression for $f_{2p}$ in \eqref{f2p} we get,

\begin{equation}\label{evenrho}
\rho_{2p} = -\frac{p}{N-1}
\end{equation}

\subsection{$\mathbf{f_{2p+1}\rho_{2p+1}}$}

\noindent We again have three cases to consider: (a) Both kernels of the upper cow-pie contracted with lower cow-pies, (b) Both kernel remain uncontracted, and (c) Only one of the kernels is contracted

\noindent Case (a) is similar to the computation of $f_{2p+1}$ and gives a factor of $(p+1)(N-p)F^{p,0,0}_{p-1,0,1;1,2}$. Case (b) does not contribute as we do not obtain the desired operator. Case (c) is similar to the case of $f_{2p}\rho_{2p}$. Once again, by explicit computation for the lowest order, we can see that its contribution is $-(p+1)F^{p,0,0}_{p-1,0,1;1,2}$.

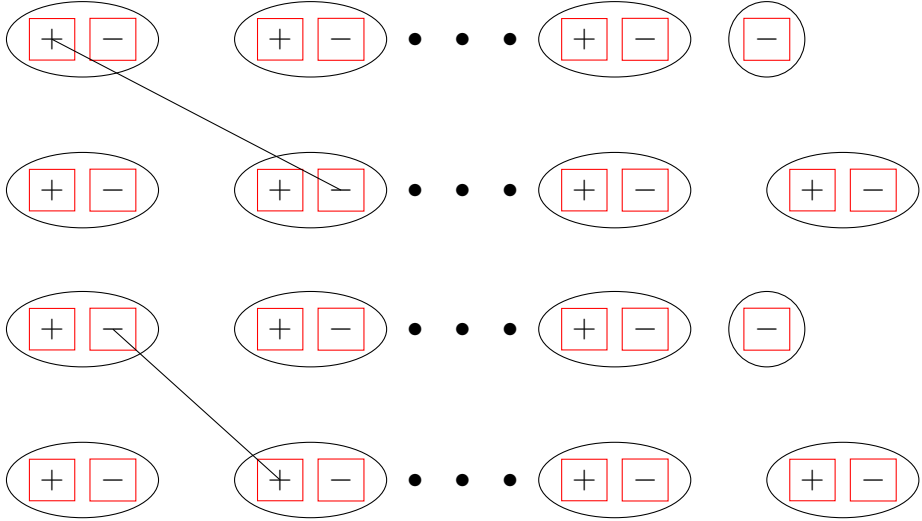
\begin{figure}[H]
\begin{tikzpicture}

\draw (5,-0.25) node[anchor=south] {\textbullet\quad\textbullet\quad\textbullet};

\begin{scope}
    \draw (0,0) ellipse (1cm and 0.5cm);
    \node[draw=red] at (-0.4,0)  {$+$};
    \node[draw=red] at (0.4,0)  {$-$} ;
\end{scope}

\begin{scope}[xshift=3cm]
\draw (0,0) ellipse (1cm and 0.5cm);
\node[draw=red] at (-0.4,0)  {$+$};
\node[draw=red] at (0.4,0)  {$-$} ;
\end{scope}
 
\begin{scope}[xshift=7cm]
\draw (0,0) ellipse (1cm and 0.5cm);
\node[draw=red] at (-0.4,0)  {$+$};
\node[draw=red] at (0.4,0)  {$-$} ;
\end{scope}

\begin{scope}[xshift=9cm]
\draw (0,0) circle (0.5cm);
\node[draw=red] at (0,0)  {$-$} ;
\end{scope}

\draw (5,-2.25) node[anchor=south] {\textbullet\quad\textbullet\quad\textbullet};

\begin{scope}[yshift=-2cm]
\draw (0,0) ellipse (1cm and 0.5cm);
\node[draw=red] at (-0.4,0)  {$+$};
\node[draw=red] at (0.4,0)  {$-$} ;
\end{scope}

\begin{scope}[xshift=3cm,yshift=-2cm]
\draw (0,0) ellipse (1cm and 0.5cm);
\node[draw=red] at (-0.4,0)  {$+$};
\node[draw=red] at (0.4,0)  {$-$} ;
\end{scope}
 
\begin{scope}[xshift=7cm,yshift=-2cm]
\draw (0,0) ellipse (1cm and 0.5cm);
\node[draw=red] at (-0.4,0)  {$+$};
\node[draw=red] at (0.4,0)  {$-$} ;
\end{scope}

\begin{scope}[xshift=10cm,yshift=-2cm]
\draw (0,0) ellipse (1cm and 0.5cm);
\node[draw=red] at (-0.4,0)  {$+$};
\node[draw=red] at (0.4,0)  {$-$} ;
\end{scope}

\draw (-0.4,0) -- (3.4,-2);


\end{tikzpicture}
\end{figure}

\begin{figure}[H]
\begin{tikzpicture}

\draw (5,-0.25) node[anchor=south] {\textbullet\quad\textbullet\quad\textbullet};

\begin{scope}
    \draw (0,0) ellipse (1cm and 0.5cm);
    \node[draw=red] at (-0.4,0)  {$+$};
    \node[draw=red] at (0.4,0)  {$-$} ;
\end{scope}

\begin{scope}[xshift=3cm]
\draw (0,0) ellipse (1cm and 0.5cm);
\node[draw=red] at (-0.4,0)  {$+$};
\node[draw=red] at (0.4,0)  {$-$} ;
\end{scope}
 
\begin{scope}[xshift=7cm]
\draw (0,0) ellipse (1cm and 0.5cm);
\node[draw=red] at (-0.4,0)  {$+$};
\node[draw=red] at (0.4,0)  {$-$} ;
\end{scope}

\begin{scope}[xshift=9cm]
\draw (0,0) circle (0.5cm);
\node[draw=red] at (0,0)  {$-$} ;
\end{scope}

\draw (5,-2.25) node[anchor=south] {\textbullet\quad\textbullet\quad\textbullet};

\begin{scope}[yshift=-2cm]
\draw (0,0) ellipse (1cm and 0.5cm);
\node[draw=red] at (-0.4,0)  {$+$};
\node[draw=red] at (0.4,0)  {$-$} ;
\end{scope}

\begin{scope}[xshift=3cm,yshift=-2cm]
\draw (0,0) ellipse (1cm and 0.5cm);
\node[draw=red] at (-0.4,0)  {$+$};
\node[draw=red] at (0.4,0)  {$-$} ;
\end{scope}
 
\begin{scope}[xshift=7cm,yshift=-2cm]
\draw (0,0) ellipse (1cm and 0.5cm);
\node[draw=red] at (-0.4,0)  {$+$};
\node[draw=red] at (0.4,0)  {$-$} ;
\end{scope}

\begin{scope}[xshift=10cm,yshift=-2cm]
\draw (0,0) ellipse (1cm and 0.5cm);
\node[draw=red] at (-0.4,0)  {$+$};
\node[draw=red] at (0.4,0)  {$-$} ;
\end{scope}

\draw (0.4,0) --  (2.6,-2);


\end{tikzpicture}
\end{figure}

\noindent So we have following recursion equation
\begin{equation}
F^{p,0,1}_{p+1,0,0;1,2} = (p+1)(N-p-1)F^{p-1,0,1}_{p,0,0;1,2}
\end{equation}

\noindent which can be solved with the launching condition $F^{1,0,0}_{0,0,1;1,2}=1$. Using \eqref{f2pp1} along with the recursion equations above, we get

\begin{equation}\label{oddrho}
\rho_{2p+1} = 1- \frac{p}{N-1}
\end{equation}

In the appendix we provide an alternate derivation of \eqref{evenrho} and \eqref{oddrho} using cow-pies.

\section{Matching with the Free Theory}

\noindent Having fixed the OPE coefficients of the free theory, we are now in a position to compute the anomalous dimensions of the interacting theory operators. This involves analyzing 3-point functions with $V_{2n}\times V^{A}_{2n+1\;a}$ OPEs in \eqref{intOPE} and demanding that in the $\e \rightarrow 0$, they go to corresponding quantities in free theory. In particular, we analyze 3-point correlators of the form

\begin{eqnarray}\label{int3pt1}
\langle V_{2n}(x_1)\;V_{2n+1\;a}^{A}(x_2)\;\bV^{B}_{1\;b}(x_3) \rangle &\rightarrow & \langle \phi_{2n}(x_1)\phi^{A}_{2n+1\;a}(x_2)\bphi^{B}_{1\;b}(x_3) \rangle \\ \nonumber  &\sim & f_{2n} (x^{2}_{12})^{-n} \langle \psi^{A}_{a}(x_2)\bpsi^{B}_{b}(x_3)
\end{eqnarray}

\noindent and

\begin{eqnarray}\label{int3pt2}
\langle V_{2n}(x_1)\;V_{2n+1\;a}^{A}(x_2)\;\bV^{B}_{3\;b}(x_3) \rangle &\rightarrow & \langle \phi_{2n}(x_1)\phi^{A}_{2n+1\;a}(x_2)\bphi^{B}_{3\;b}(x_3) \rangle \\ \nonumber & \sim & f_{2n}\rho_{2n} (x^{2}_{12})^{-n} (\fsl{x}_{12})_{ab} \langle \phi^{A}_{3\;a}(x_2)\bphi^{B}_{3\;b}(x_3)\rangle
\end{eqnarray}

\noindent The LHS of \eqref{int3pt2} can be evaluated, to the leading order, using $V_{2n}\times V^{A}_{2n+1\;a}$ OPE of \eqref{intOPE} and the fact that $V_{3}$ is a descendent of $V_1$, i.e,

\bea
\langle V^{A}_{1\;a}(x_1)\bV^{B}_{3\;b} \rangle = \alpha^{-1}(\e) \partial_{2\;\mu} \langle V^{A}_{1\;a}(x_1) \bV^{B}_{1\;c} \rangle (\gamma^{\mu})_{cb} = \sigma\sqrt{(N-1)\gamma_1}\frac{\delta_{ab} \delta^{AB}}{(x^{2}_{12})^{\Delta_1 + \frac{1}{2}}}
\eea

\noindent Since this is proportional to $\sqrt{\gamma_1}$, it vanishes in the $\e \rightarrow 0$ limit. Therefore, to reproduce \eqref{freeOPE1} and \eqref{freeOPE2} we need $q_1$ to remain finite in this limit. We also need $q_2$ to blow up as $\e \rightarrow 0$ limit such that 

\bea
q^{i}_2 \alpha(\e) \rightarrow \rho_{i}, \quad i=2p,\;2p+1
\eea

\noindent The coefficients $q_i$ are determined by conformal symmetry whose details and explicit form can be found in Appendix. As alluded before, we find that $q_1$ is indeed finite in the free limit. The asymptotic behavior of $q_2$ is given by

\bea
q_{2}^{2n}\approx \frac{(\gamma_1 + \gamma_{2n}-\gamma_{2n+1})}{4 \gamma_1}, \qquad q_{2}^{2n+1}\approx \frac{(\gamma_1 + \gamma_{2n+1}-\gamma_{2n+2})}{4 \gamma_1}
\eea

\noindent Its evident that for $q_2$ to blow up $y_{1,1}$ has to vanish. This gives us following telescoping series

\begin{eqnarray}
y_{2n,1}-y_{2n+1,1} &=& 2\sigma \sqrt{(N-1)y_{1,2}}\;\rho_{2n}, \quad n = 1,2,\cdots \\ 
y_{2n+1,1}-y_{2n+2,1} &=& 2\sigma \sqrt{(N-1)y_{1,2}}\;\rho_{2n+1} \quad n = 0,1,\cdots
\end{eqnarray}

\noindent Together this can be written as

\bea
y_{i,1} - y_{i+1,1} = 2\sigma \sqrt{(N-1)y_{1,2}}\;\rho_{i}, \qquad i=1,2,\cdots
\eea

\noindent Summing the telescoping series gives

\begin{eqnarray}\label{anomdim}
y_{n,1} = K\sum^{n-1}_{m=1}\rho_{m}
\end{eqnarray}

\noindent This gives the anomalous dimensions of all the odd and even primaries in the theory once we fix the numerical value of $K$. To fix this we make use of \eqref{dimconstraint} which can be written as

\bea
2\delta + \gamma_3 = \gamma_1 + 1
\eea

\noindent This gives $y_{3,1}=-1$ which can now be used to fix $K$ by setting $n=3$ in \eqref{anomdim}.

\begin{equation}
y_{3,1} = K(\rho_1 + \rho_2)
\end{equation}

\noindent This gives $K=-\frac{(N-1)}{(N-2)}$ which fixes $\sigma=-1$ and furthermore fixes $y_{1,2}$ also. Thus we obtain

\begin{equation}\label{gamma1}
\gamma_1 = \frac{(N-1)}{4(N-2)^2}\e^2
\end{equation}

\noindent One can also compute the anomalous dimensions of $V_2$ which we obtain to be

\begin{equation}\label{gamma2}
\gamma_2 = -\frac{(N-1)}{(N-2)}\e
\end{equation}

\noindent which are in perfect agreement with the results of \cite{Derkachov,Gracey1,Gracey2}.

\section*{Acknowledgments}
\noindent I thank Chethan Krishnan for suggesting the problem, collaboration at various stages and several useful discussions, without which this paper would not have seen the daylight.


\appendix

\section{OPE coefficients from 3-point function}

As mentioned in Section \ref{sec1}, the OPE coefficients, $q_i$, are completely determined by the conformal symmetry \cite{Ferrara}. Here we outline a procedure for obtaining these coefficients from an expansion of 3-point functions. For the case in hand, the coefficients are computed from a scalar-fermion-antifermion 3-pt correlator which takes following form \cite{Weinberg} 

\bea\label{gen3ptfun}
\langle V_{2n}(x_1)V^{A}_{2n+1\;a}(x_2)\bV^{B}_{1\;b}(x_3) \rangle = C_{123} \frac{(\fsl{x}_{23})_{ab}\delta^{AB}}{(x^{2}_{12})^{l_3}\;(x^{2}_{23})^{l_1}\;(x^{2}_{31})^{l_2}}
\eea 

\noindent where $l_1$, $l_2$ and $l_3$ is determined in terms of the scaling dimensions of the operators

\begin{eqnarray}
l_1 &=& \frac{1}{2}\left[1-\Delta_{2n}+\Delta_{2n+1}+\Delta_1 \right] \\ \nonumber
l_2 &=& \frac{1}{2}\left[\Delta_{2n}-\Delta_{2n+1}+\Delta_1 \right] \\ \nonumber
l_3 &=& \frac{1}{2}\left[\Delta_{2n}+\Delta_{2n+1}-\Delta_1 \right] \\ 
\end{eqnarray}

\noindent Now we imagine a scenario where the first two operators, $V_{2n}(x_1)$ and $V^{A}_{2n+1\;a}(x_2)$, are coming together such that $|x_{12}|\ll|x_{31}|$ and $|x_{12}|\ll|x_{23}|$.  This allows us to expand the 3-pt function \eqref{gen3ptfun} by eliminating $x_{31}$ using the relation

\bea
x_{31}^{2} = x^{2}_{23}\left( 1 + \frac{2 x_{12}.x_{23}}{x^{2}_{23}} + \frac{x^{2}_{12}}{x^{2}_{23}} \right)
\eea

\noindent Substituting this in \eqref{gen3ptfun} and keeping terms upto $O(x_{12})$ we obtain following series

\begin{eqnarray}
&\;& \langle V_{2n}(x_1)V^{A}_{2n+1\;a}(x_2)\bV^{B}_{1\;b}(x_3) \rangle \equiv C_{123} \frac{(\fsl{x}_{23})_{ab}\delta^{AB}}{(x^{2}_{12})^{l_3}\;(x^{2}_{23})^{l_1}\;(x^{2}_{31})^{l_2}} \\ \nonumber &\approx & C_{123} (x^{2}_{12})^{-\frac{1}{2}\left[ \Delta_{2n} + \Delta_{2n+1}-\Delta_1 \right]} \left[ \frac{(\fsl{x}_{23})_{ab}}{(x^{2}_{23})^{\Delta_1 + \frac{1}{2}}} - 2l_2 \frac{\fsl{x}_{23})_{ab}(x_{12}.x_{23})}{(x^{2}_{23})^{\Delta_1 + \frac{3}{2}}}  \right]
\end{eqnarray}

\noindent Since the operators $V_{2n}$ and $V^{A}_{2n+1\;a}(x_2)$ are close, we may use OPE \eqref{intOPE}. Substituting this into the LHS of \eqref{gen3ptfun}, we obtain

\begin{eqnarray}
& \;& \langle  V_{2n}(x_1) V^{A}_{2n+1\;a}(x_2)\bV^{B}_{1\;b}(x_3) \rangle \approx \tilde{f}(x^{2}_{12})^{-\frac{1}{2}[\Delta_{2n}+\Delta_{2n+1}-\Delta_1]} \Big[ \langle V^{A}_{1\;a}(x_2)\bV^{B}_{1\;b}(x_3) \rangle \\ \nonumber  &+& q_1\; x_{12}^{\mu}\partial_{2\; \mu} \langle V^{A}_{1\;a}(x_2)\bV^{B}_{1\;b}(x_3)\rangle + q_2\;(\fsl{x}_{12}\fsl{\partial}_{2})_{ac}\langle V^{A}_{1\;c}(x_2)\bV^{B}_{1\;b}(x_3) \rangle \Big]
\end{eqnarray}

\noindent This evaluates to

\begin{eqnarray}
& \;& \langle  V_{2n}(x_1) V^{A}_{2n+1\;a}(x_2)\bV^{B}_{1\;b}(x_3) \rangle \approx \tilde{f}(x^{2}_{12})^{-\frac{1}{2}[\Delta_{2n}+\Delta_{2n+1}-\Delta_1]} \left[ \frac{(\fsl{x}_{23})_{ab}}{(x^{2}_{23})^{\Delta_1 + \frac{1}{2}}} \right. \\ \nonumber &+& \left. q_1 \left( \frac{(\fsl{x}_{23})_{ab}}{(x^{2}_{12})^{\Delta_1 + \frac{1}{2}}}  - \frac{(2\Delta_1 + 1)(x_{12}.x_{23})(\fsl{x}_{23})_{ab}}{(x^{2}_{23})^{\Delta_1 + \frac{3}{2}}} \right) + q_2\;\frac{(d-2\Delta_1 -1)(\fsl{x}_{12})_{ab}}{(x^{2}_{23})^{\Delta_1 + \frac{1}{2}}}  \right]\delta^{AB}
\end{eqnarray}

\noindent Comparing this with the 3-pt expansion \eqref{gen3ptfun}, we get

\begin{eqnarray}\label{coeff1}
q_1 &=& \frac{\Delta_{2n}-\Delta_{2n+1}+\Delta_1}{2\Delta_1 + 1} \\ \nonumber
q_2 &=& \frac{\Delta_{2n}-\Delta_{2n+1}+\Delta_1}{(2\Delta_1 + 1)(2\Delta_1 + 1 - d)} \\ \nonumber
\end{eqnarray}

\noindent Similar story holds for the fermion-scalar-anti-fermion 3-point function also. Here the 3-point function is given by

\begin{eqnarray}\label{gen3ptfun2}
\langle V^{A}_{2n+1\;a}(x_1)V_{2n+2}(x_2)\bV^{B}_{1\;b}(x_3) \rangle = C'_{123} \frac{(\fsl{x}_{23})_{ab}\delta^{AB}}{(x^{2}_{12})^{m_3}\;(x^{2}_{23})^{m_1}\;(x^{2}_{31})^{m_2}} 
\end{eqnarray}

\noindent with

\begin{eqnarray}
m_1 &=& \frac{1}{2}\left[1-\Delta_{2n+1}+\Delta_{2n+2}+\Delta_1 \right] \\ \nonumber
m_2 &=& \frac{1}{2}\left[\Delta_{2n+1}-\Delta_{2n+2}+\Delta_1 \right] \\ \nonumber
m_3 &=& \frac{1}{2}\left[1+\Delta_{2n+1}+\Delta_{2n+2}-\Delta_1 \right] \nonumber
\end{eqnarray}

\noindent Proceeding in a similar manner, the 3-point function expansion takes the form

\begin{eqnarray}\label{3pt2exp}
&\;& \langle V^{A}_{2n+1\;a}(x_1)V_{2n+2}(x_2)\bV^{B}_{1\;b}(x_3) \rangle \equiv C'_{123} \frac{(\fsl{x}_{23})_{ab}\delta^{AB}}{(x^{2}_{12})^{m_3}\;(x^{2}_{23})^{m_1}\;(x^{2}_{31})^{m_2}} \\ \nonumber &\approx & C'_{123} (x^{2}_{12})^{-\frac{1}{2}\left[ \Delta_{2n+1}+\Delta_{2n+2}-\Delta_1 + 1 \right]}(\fsl{x}_{12})_{ac}\left[ \frac{(\fsl{x}_{23})_{ab}}{(x^{2}_{23})^{\Delta_1 + \frac{1}{2}}} - 2l_2 \frac{\fsl{x}_{23})_{ab}(x_{12}.x_{23})}{(x^{2}_{23})^{\Delta_1 + \frac{3}{2}}}  \right]
\end{eqnarray}

\noindent On the other hand, OPE of the first two operators is given by

\begin{eqnarray}
V^{A}_{2n+1\;a}(x_1)\times V_{2n+2}(x_2) & \approx & \tilde{f}(x^{2}_{12})^{-\frac{1}{2}\left[ \Delta_{2n+1}+\Delta_{2n+2}-\Delta_1 + 1 \right]} (\fsl{x}_{12})_{ac} \\ \nonumber &\times & \Big[ \delta_{cd} + q_1\;\delta_{cd}x^{\mu}_{12}\partial_{2\;\mu} + q_2\;(\fsl{x}_{12}\fsl{\partial}_{2})_{cd} \Big]V^{A}_{1\;d}(x_2)
\end{eqnarray}

\noindent Substituting this in the LHS of \eqref{gen3ptfun2} and comparing with \eqref{3pt2exp}, we get

\begin{eqnarray}
q_1 &=& \frac{\Delta_{2n+1}-\Delta_{2n+2}+\Delta_1}{2\Delta_1 + 1} \\ \nonumber
q_2 &=& \frac{\Delta_{2n+1}-\Delta_{2n+2}+\Delta_1}{(2\Delta_1 + 1)(2\Delta_1 + 1 - d)} \\ \nonumber 
\end{eqnarray}

\section{Computing $f_{2p}\rho_{2p}$ and $f_{2p+1}\rho_{2p+1}$ from cow-pies}

In this appendix we give an alternate way to obtain $f_{2p}\rho_{2p}$ and $f_{2p+1}\rho_{2p+1}$ coefficients using cow-pie contractions. This works as a double check of our results, because there are not many results other than \eqref{gamma1} and \eqref{gamma2} that we can check in the literature.

\subsection{$\mathbf{f_{2p}\rho_{2p}}$}

\noindent For the ease of counting, we invert the cow-pie and start the contractions from the single kernel. There are two cases to consider,

\begin{enumerate}
\item[\textbf{Case I}]: The `$-$' kernel remains uncontracted. It is easy to see that its contribution is zero because rest of the contractions cannot give the desired operator.

\item[\textbf{Case II}]: The `$-$' kernel is contracted with `$+$' from the double cow-pie.

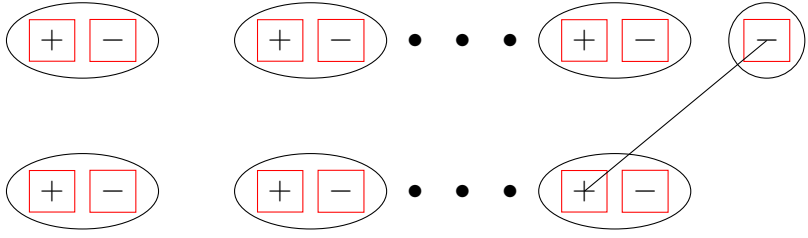
\begin{figure}[H]
\begin{tikzpicture}

\draw (5,-0.25) node[anchor=south] {\textbullet\quad\textbullet\quad\textbullet};

\begin{scope}
    \draw (0,0) ellipse (1cm and 0.5cm);
    \node[draw=red] at (-0.4,0)  {$+$};
    \node[draw=red] at (0.4,0)  {$-$} ;
\end{scope}

\begin{scope}[xshift=3cm]
\draw (0,0) ellipse (1cm and 0.5cm);
\node[draw=red] at (-0.4,0)  {$+$};
\node[draw=red] at (0.4,0)  {$-$} ;
\end{scope}
 
\begin{scope}[xshift=7cm]
\draw (0,0) ellipse (1cm and 0.5cm);
\node[draw=red] at (-0.4,0)  {$+$};
\node[draw=red] at (0.4,0)  {$-$} ;
\end{scope}

\begin{scope}[xshift=9cm]
\draw (0,0) circle (0.5cm);
\node[draw=red] at (0,0)  {$-$} ;
\end{scope}

\draw (5,-2.25) node[anchor=south] {\textbullet\quad\textbullet\quad\textbullet};

\begin{scope}[yshift=-2cm]
\draw (0,0) ellipse (1cm and 0.5cm);
\node[draw=red] at (-0.4,0)  {$+$};
\node[draw=red] at (0.4,0)  {$-$} ;
\end{scope}

\begin{scope}[xshift=3cm,yshift=-2cm]
\draw (0,0) ellipse (1cm and 0.5cm);
\node[draw=red] at (-0.4,0)  {$+$};
\node[draw=red] at (0.4,0)  {$-$} ;
\end{scope}
 
\begin{scope}[xshift=7cm,yshift=-2cm]
\draw (0,0) ellipse (1cm and 0.5cm);
\node[draw=red] at (-0.4,0)  {$+$};
\node[draw=red] at (0.4,0)  {$-$} ;
\end{scope}

\draw (9.0,0) -- (6.6,-2);


\end{tikzpicture}
\end{figure}

which contributes $-p F^{p-1,0,1}_{p,0,0;1,1}$ which can again be evaluated using cow-pies. Once again, we invert the cow-pie diagram and start the contractions from the uncontracted `$-$' in the double cow-pie. As can be readily seen, there are two cases to consider:

\begin{enumerate}
\item[\textbf{(a)}]: `$-$' remains uncontracted. This gives a contribution of $F^{p,0,0}_{p+1,0,0;1,1}$. 

\begin{figure}[H]
\begin{tikzpicture}

\draw (5,-0.25) node[anchor=south] {\textbullet\quad\textbullet\quad\textbullet};

\begin{scope}
    \draw (0,0) ellipse (1cm and 0.5cm);
    \node[draw=red] at (-0.4,0)  {$+$};
    \node[draw=red] at (0.4,0)  {$-$} ;
\end{scope}

\begin{scope}[xshift=3cm]
\draw (0,0) ellipse (1cm and 0.5cm);
\node[draw=red] at (-0.4,0)  {$+$};
\node[draw=red] at (0.4,0)  {$-$} ;
\end{scope}
 
\begin{scope}[xshift=7cm]
\draw (0,0) ellipse (1cm and 0.5cm);
\node[draw=red] at (-0.4,0)  {$+$};
\node[draw=red] at (0.4,0)  {$-$} ;
\end{scope}

\draw (5,-2.25) node[anchor=south] {\textbullet\quad\textbullet\quad\textbullet};

\begin{scope}[yshift=-2cm]
\draw (0,0) ellipse (1cm and 0.5cm);
\node[draw=red] at (-0.4,0)  {$+$};
\node[draw=red] at (0.4,0)  {$-$} ;
\end{scope}

\begin{scope}[xshift=3cm,yshift=-2cm]
\draw (0,0) ellipse (1cm and 0.5cm);
\node[draw=red] at (-0.4,0)  {$+$};
\node[draw=red] at (0.4,0)  {$-$} ;
\end{scope}
 
\begin{scope}[xshift=7cm,yshift=-2cm]
\draw (0,0) ellipse (1cm and 0.5cm);
\node[draw=red] at (-0.4,0)  {$+$};
\node[draw=red] at (0.4,0)  {$-$} ;
\end{scope}

\begin{scope}[xshift=9cm,yshift=-2cm]
\draw (0,0) circle (0.5cm);
\node[draw=red] at (0,0)  {$-$} ;
\end{scope}

\draw (6.6,0) -- (9.0,-2);


\end{tikzpicture}
\end{figure}

$F^{p,0,0}_{p+1,0,0;1,1}$ can in turn be evaluated using cow-pies. This is similar to the case of $f_{2p}$ and $f_{2p+1}$ and its recursion equation is given by

\begin{equation}\label{recursion2}
F^{p,0,0}_{p+1,0,0;1,1} = (p+1)(N-p)F^{p-1,0,0}_{p,0,0;1,1}
\end{equation}

\item[\textbf{(b)}]: `$-$' can be contracted with one of the double cow-pies. This gives a factor of $(p+1)F^{p,0,0}_{p,0,1;1,2}$

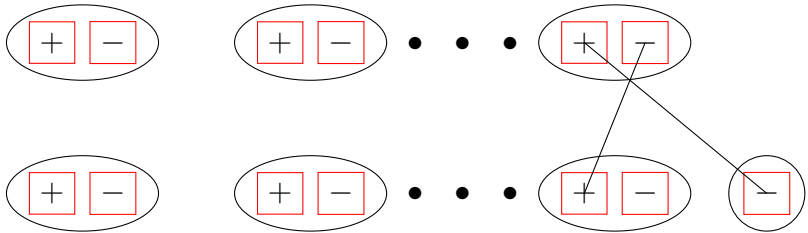
\begin{figure}[H]
\begin{tikzpicture}

\draw (5,-0.25) node[anchor=south] {\textbullet\quad\textbullet\quad\textbullet};

\begin{scope}
    \draw (0,0) ellipse (1cm and 0.5cm);
    \node[draw=red] at (-0.4,0)  {$+$};
    \node[draw=red] at (0.4,0)  {$-$} ;
\end{scope}

\begin{scope}[xshift=3cm]
\draw (0,0) ellipse (1cm and 0.5cm);
\node[draw=red] at (-0.4,0)  {$+$};
\node[draw=red] at (0.4,0)  {$-$} ;
\end{scope}
 
\begin{scope}[xshift=7cm]
\draw (0,0) ellipse (1cm and 0.5cm);
\node[draw=red] at (-0.4,0)  {$+$};
\node[draw=red] at (0.4,0)  {$-$} ;
\end{scope}

\draw (5,-2.25) node[anchor=south] {\textbullet\quad\textbullet\quad\textbullet};

\begin{scope}[yshift=-2cm]
\draw (0,0) ellipse (1cm and 0.5cm);
\node[draw=red] at (-0.4,0)  {$+$};
\node[draw=red] at (0.4,0)  {$-$} ;
\end{scope}

\begin{scope}[xshift=3cm,yshift=-2cm]
\draw (0,0) ellipse (1cm and 0.5cm);
\node[draw=red] at (-0.4,0)  {$+$};
\node[draw=red] at (0.4,0)  {$-$} ;
\end{scope}
 
\begin{scope}[xshift=7cm,yshift=-2cm]
\draw (0,0) ellipse (1cm and 0.5cm);
\node[draw=red] at (-0.4,0)  {$+$};
\node[draw=red] at (0.4,0)  {$-$} ;
\end{scope}

\begin{scope}[xshift=9cm,yshift=-2cm]
\draw (0,0) circle (0.5cm);
\node[draw=red] at (0,0)  {$-$} ;
\end{scope}

\draw (6.6,0) -- (9.0,-2);
\draw (7.4,0) --  (6.6,-2);


\end{tikzpicture}
\end{figure}

\end{enumerate}
\end{enumerate}

\noindent Putting all the pieces together, we have the following system of recursion equations

\begin{eqnarray}
F^{p,0,0}_{p,0,1;1,2} &=& -p F^{p-1,0,1}_{p,0,0;1,1} \\ \nonumber
F^{p,0,1}_{p+1,0,0;1,1} &=& F^{p,0,0}_{p+1,0,0;1,1} + (p+1)F^{p,0,0}_{p,0,1;1,2} \\ \nonumber
F^{p,0,0}_{p+1,0,0;1,1} &=& (p+1)(N-p)F^{p-1,0,0}_{p,0,0;1,1} \nonumber
\end{eqnarray}

\noindent which can be solved along with the launching conditions $F^{0,0,0}_{0,0,1;1,2}=0$, $F^{0,0,1}_{1,0,0;1,1}=1$ and $F^{0,0,0}_{1,0,0;1,1}=1$. Using the expression for $f_{2p}$ in \eqref{f2p}, we obtain

\begin{equation}
\rho_{2p} = - \frac{p}{N-1}
\end{equation}

\subsection{$\mathbf{f_{2p+1}\rho_{2p+1}}$}

\noindent We proceed analogous to the even case, \textit{i.e.} $f_{2p}\rho_{2p}$. As in the previous case, we start the contractions with the single kernel `$-$'. Again, we have two cases:

\begin{enumerate}
\item[\textbf{Case I}]: `$-$' remains uncontracted. This indeed contributes, with a factor of $F^{p,0,0}_{p+1,0,0;1,1}$ which can be further evaluated using cow-pies. It can be seen that the recursion equation for $F^{p,0,0}_{p+1,0,0;1,1}$ is given by

\begin{equation}
F^{p,0,0}_{p+1,0,0;1,1} = (p+1)(N-p)F^{p-1,0,0}_{p,0,0;1,1}
\end{equation}

\item[\textbf{Case II}]: `$-$' is contracted with one of the double cow-pies. This contributes a factor of $(p+1)F^{p,0,0}_{p,0,1;1,2}$, where $F^{p,0,0}_{p,0,1;1,2}$ can be furthermore evaluated using cow-pies. To this end, we invert the cow-pie diagram and once again consider two separate cases 

\begin{enumerate}
\item[\textbf{a}]: `$-$' of the double cow-pie remains uncontracted. It can be seen that this does not contribute as we do not get the desired operator.

\item[\textbf{b}]: `$-$' of the double cow-pie is contracted with one of the double cow-pies. This contributes $-pF^{p-1,0,1}_{p,0,0;1,2}$.

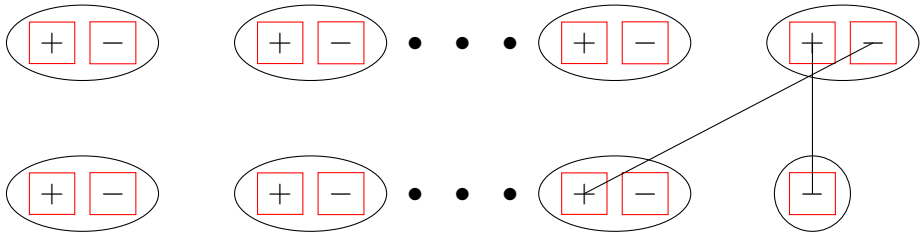
\begin{figure}[H]
\begin{tikzpicture}

\draw (5,-0.25) node[anchor=south] {\textbullet\quad\textbullet\quad\textbullet};

\begin{scope}
    \draw (0,0) ellipse (1cm and 0.5cm);
    \node[draw=red] at (-0.4,0)  {$+$};
    \node[draw=red] at (0.4,0)  {$-$} ;
\end{scope}

\begin{scope}[xshift=3cm]
\draw (0,0) ellipse (1cm and 0.5cm);
\node[draw=red] at (-0.4,0)  {$+$};
\node[draw=red] at (0.4,0)  {$-$} ;
\end{scope}
 
\begin{scope}[xshift=7cm]
\draw (0,0) ellipse (1cm and 0.5cm);
\node[draw=red] at (-0.4,0)  {$+$};
\node[draw=red] at (0.4,0)  {$-$} ;
\end{scope}

\begin{scope}[xshift=10cm]
\draw (0,0) ellipse (1cm and 0.5cm);
\node[draw=red] at (-0.4,0)  {$+$};
\node[draw=red] at (0.4,0)  {$-$} ;
\end{scope}


\draw (5,-2.25) node[anchor=south] {\textbullet\quad\textbullet\quad\textbullet};

\begin{scope}[yshift=-2cm]
\draw (0,0) ellipse (1cm and 0.5cm);
\node[draw=red] at (-0.4,0)  {$+$};
\node[draw=red] at (0.4,0)  {$-$} ;
\end{scope}

\begin{scope}[xshift=3cm,yshift=-2cm]
\draw (0,0) ellipse (1cm and 0.5cm);
\node[draw=red] at (-0.4,0)  {$+$};
\node[draw=red] at (0.4,0)  {$-$} ;
\end{scope}
 
\begin{scope}[xshift=7cm,yshift=-2cm]
\draw (0,0) ellipse (1cm and 0.5cm);
\node[draw=red] at (-0.4,0)  {$+$};
\node[draw=red] at (0.4,0)  {$-$} ;
\end{scope}

\begin{scope}[xshift=9.6cm,yshift=-2cm]
\draw (0,0) circle (0.5cm);
\node[draw=red] at (0,0)  {$-$} ;
\end{scope}

\draw (9.6,0) -- (9.6,-2);
\draw (10.4,0) --  (6.6,-2);


\end{tikzpicture}
\end{figure}
\end{enumerate}
\end{enumerate}

\noindent Thus we have following system of recursion equations, which can be solved along with the launching conditions $F{0,0,1}_{1,0,0;1,2}=1$, $F^{0,0,0}_{1,0,0;1,1}=1$, $F^{0,0,0}_{0,0,1;1,2}=0$.

\begin{eqnarray}
F^{p,0,1}_{p+1,0,0;1,2} &=& F^{p,0,0}_{p+1,0,0;1,1} + (p+1)F^{p,0,0}_{p,0,1;1,2} \\ \nonumber
F^{p,0,0}_{p+1,0,0;1,1} &=& (p+1)(N-p)F^{p-1,0,0}_{p,0,0;1,1}  \\ \nonumber
F^{p,0,0}_{p,0,1;1,2} &=& -p F^{p-1,0,1}_{p,0,1;1,2} \\ \nonumber
\end{eqnarray} 

\noindent Using \eqref{f2pp1}, along with above set of recursion equations gives

\begin{equation}
\rho_{2p+1} = 1- \frac{p}{N-1}
\end{equation}


\begin{thebibliography}{99}
\bibitem{Rychkov} 
  S.~Rychkov and Z.~M.~Tan,
  {\em The Epsilon-Expansion from Conformal Field Theory},
  arXiv:1505.00963 [hep-th].
  
\bibitem{Chethan-Pallab} 
  P.~Basu and C.~Krishnan,
  {\em $\epsilon$-Expansions Near Three Dimensions from Conformal Field Theory},
  arXiv:1506.06616 [hep-th].

\bibitem{GN} 
  D.~J.~Gross and A.~Neveu,
  {\em Dynamical Symmetry Breaking in Asymptotically Free Field Theories},
  Phys.\ Rev.\ D {\bf 10}, 3235 (1974).
  
\bibitem{Vasiliev1} 
  A.~N.~Vasiliev and M.~I.~Vyazovsky,
  {\em Proof of the absence of multiplicative renormalizability of the Gross-Neveu model in the dimensional regularization d = 2+2epsilon},
  Theor.\ Math.\ Phys.\  {\bf 113}, 1277 (1997)
  [Teor.\ Mat.\ Fiz.\  {\bf 113}, 85 (1997)].
  
\bibitem{Vasiliev2} 
  A.~N.~Vasiliev, M.~I.~Vyazovsky, S.~E.~Derkachov and N.~A.~Kivel,
  {\em On the equivalence of renormalizations in standard and dimensional regularizations of 2-D four-fermion interactions},
  Theor.\ Math.\ Phys.\  {\bf 107}, 441 (1996)
  [Teor.\ Mat.\ Fiz.\  {\bf 107}, 27 (1996)].


\bibitem{Derkachov} 
  S.~E.~Derkachov, N.~A.~Kivel, A.~S.~Stepanenko and A.~N.~Vasiliev,
  {\em On calculation in 1/n expansions of critical exponents in the Gross-Neveu model with the conformal technique},
  hep-th/9302034.
  
\bibitem{Gracey1} 
  J.~A.~Gracey,
  {\em Computation of the three loop Beta function of the O(N) Gross-Neveu model in minimal subtraction},
  Nucl.\ Phys.\ B {\bf 367}, 657 (1991).
  
\bibitem{Gracey2} 
  J.~A.~Gracey,
  {\em Three loop calculations in the O(N) Gross-Neveu model},
  Nucl.\ Phys.\ B {\bf 341}, 403 (1990).
  
\bibitem{Ferrara} 
  S.~Ferrara, R.~Gatto and A.~F.~Grillo,
  {\em Conformal algebra in space-time and operator product expansion},
  Springer Tracts Mod.\ Phys.\  {\bf 67}, 1 (1973).
  
  
\bibitem{Weinberg} 
  S.~Weinberg,
  {\em Six-dimensional Methods for Four-dimensional Conformal Field Theories},
  Phys.\ Rev.\ D {\bf 82}, 045031 (2010)
  [arXiv:1006.3480 [hep-th]].

\bibitem{rgupta} 
  S.~Ghosh, R.~K.~Gupta, K.~Jaswin and A.~A.~Nizami,
  {\em $\epsilon$-Expansion in the Gross-Neveu Model from Conformal Field Theory},
  arXiv:1510.04887 [hep-th].
  
\end{thebibliography}
\end{document}